\documentclass[conference]{IEEEtran}

\usepackage{color}
\usepackage{amsmath,amssymb,amsfonts}
\usepackage{graphicx}
\usepackage{verbatim}

\usepackage[font=scriptsize,caption=false,labelsep=space]{subfig}

\usepackage{url}
\usepackage[para,online,flushleft]{threeparttable}

\usepackage{booktabs}
\usepackage{multicol}

\usepackage{enumitem}

\usepackage[switch]{lineno}
\usepackage[named]{algo}
\usepackage[noend]{algpseudocode}
\usepackage{algorithm}

\usepackage{cite}
\makeatletter
\newcommand{\citecomment}[2][]{\citen{#2}#1\citevar}
\newcommand{\citeone}[1]{\citecomment{#1}}
\newcommand{\citetwo}[2][]{\citecomment[,~#1]{#2}}
\newcommand{\citevar}{\@ifnextchar\bgroup{;~\citeone}{\@ifnextchar[{;~\citetwo}{]}}}
\newcommand{\citefirst}{\@ifnextchar\bgroup{\citeone}{\@ifnextchar[{\citetwo}{]}}}

\makeatother

\usepackage{balance}

\usepackage{amsmath}
\usepackage{amsthm}
\usepackage{dsfont}
\usepackage{thmtools}
\usepackage{amssymb}
\usepackage[dvipsnames]{xcolor}

\def\bn{\boldsymbol{n}}

\def\bGamma{\boldsymbol{\Gamma}}

\def\btheta{\boldsymbol{\theta}}

\def\bmu{\boldsymbol{\mu}}

\def\mbe{\mathbf{e}}
\def\mbf{\mathbf{f}}
\def\mbg{\mathbf{g}}
\def\mbh{\mathbf{h}}

\def\mbn{\mathbf{n}}

\def\mbs{\mathbf{s}}

\def\mbx{\mathbf{x}}
\def\mby{\mathbf{y}}

\def\mbA{\mathbf{A}}

\def\mbG{\mathbf{G}}
\def\mbH{\mathbf{H}}
\def\mbI{\mathbf{I}}

\def\mbP{\mathbf{P}}

\def\mbW{\mathbf{W}}

\def\bzero{\boldsymbol{0}}
\def\bone{\boldsymbol{1}}

\newcommand{\expecE}[1]{\mathds{E}\left\{{#1}\right\}}

\newcommand{\realR}[1]{\mathds{R}^{#1}}

\theoremstyle{definition}

\def\T{\top}

\makeatletter
\newcommand*{\rom}[1]{\expandafter\@slowromancap\romannumeral #1@}
\makeatother

\usepackage[english]{babel}

\begin{document}
\title{Deep Learning Meets Adaptive Filtering: \\ A Stein's Unbiased Risk Estimator Approach}
\author{Zahra Esmaeilbeig and Mojtaba Soltanalian}
\author{\IEEEauthorblockN{Zahra Esmaeilbeig, Mojtaba Soltanalian}
		\IEEEauthorblockA{ WaveOPT Lab, 
			University of Illinois Chicago, USA\\
			\IEEEauthorblockA{
				Email: \{zesmae2, msol\}@uic.edu}
		}
	}
\maketitle

\begin{abstract}
This paper revisits two prominent adaptive filtering algorithms, namely recursive least squares (RLS) and equivariant adaptive source separation (EASI), through the lens of algorithm unrolling. Building upon the unrolling methodology, we introduce novel task-based deep learning frameworks, denoted as \textit{Deep RLS} and \textit{Deep EASI}. These architectures transform the iterations of the original algorithms into layers of a deep neural network, enabling efficient source signal estimation by leveraging a training process. To further enhance performance, we propose training these deep unrolled networks utilizing a  surrogate loss function grounded on  Stein's unbiased risk estimator (SURE). Our empirical evaluations demonstrate that the \textit{Deep RLS} and \textit{Deep EASI} networks outperform their underlying algorithms. Moreover,  the efficacy of SURE-based training in comparison to  conventional  mean squared error loss is highlighted by numerical experiments. The unleashed potential of SURE-based training in this paper sets a benchmark for future employment of SURE  either for  training purposes or as an evaluation metric for generalization performance of neural networks. 
\end{abstract}
\IEEEpeerreviewmaketitle
\section{Introduction}

\IEEEPARstart{}{} 
 \textit{Deep unfolding}, or \textit{unrolling}~\cite{hershey2014deep}, enables  constructing interpretable  deep neural networks (DNN) that require less training data and considerably less computational resources than generic DNNs.
Specifically, in deep unfolding, each layer of the DNN is designed to resemble one
iteration of the original algorithm of interest. Passing the signals through such a deep network is in essence similar to executing
the iterative algorithm a finite number of times, determined by the number of layers. The 
model parameters  will be reflected in weights of the constructed DNN.
 The data-driven nature of the emerging deep network thus enables improvements over the original algorithm.  Note that the constructed network may be trained using back-propagation, resulting
in model parameters that are learned from the real-world training
datasets. 
In this way, the trained network can be naturally
interpreted as a parameter optimized algorithm, effectively
overcoming the lack of interpretability in most conventional
neural networks~\cite{monga2019algorithm}.
In comparison with a generic DNN, the unfolded network has many fewer parameters and therefore requires a more modest size of training data and  computational resources. 

The deep unrolling technique has been effectively applied to various signal processing problems, yielding significant improvements in the convergence rates of state-of-the-art iterative algorithms; see~\cite{monga2019algorithm,hershey2014deep} for a detailed explanation of deep unrolling, as well as~\cite{shlezinger2022model,zeng2022one,khobahi2021lord} for examples of deploying deep unrolling in different application areas.

Our goal in this  paper is  to develop a set of algorithms able to  learn the nonlinearity and  step sizes  of  two classical adaptive  filtering algorithms, namely, recursive least squares (RLS) and  equivariant adaptive source separation (EASI).  
We leverage Stein’s unbiased risk
estimator (SURE) in  training, which serves as a surrogate for mean squared error (MSE), even when the
ground truth is unknown~\cite{Edupuganti2021}. Studies such as~\cite{metzler2018unsupervised,mardani2019degrees} have reported improved image denoising results when networks were trained with SURE loss, outperforming traditional MSE loss training. Similarly, SURE has been effectively used to train deep convolutional neural networks without requiring denoised ground truth, as highlighted in~\cite{metzler2018unsupervised,shamshad2018leveraging}.
The SURE based training and the recurrent  training  procedure of  our  proposed methodology makes it a great candidate for unsupervised real-time signal processing.
 

The rest of the paper is organized in the following manner. In section II, we introduce the problem formulation for adaptive filtering-based signal estimation. In Section~\ref{sec::framework}, we propose  the  two  deep unrolling frameworks \textit{Deep EASI} and \textit{Deep RLS}  for  adaptive  filtering, alongside the SURE-based surrogate loss function employed for their training. Section~\ref{sec::numerical} details the numerical experiments used to evaluate our proposed methods, and Section~\ref{sec:conclusion} presents our concluding remarks.

\emph{Notation:} Throughout this paper, we use bold lowercase and bold uppercase letters for vectors and matrices, respectively. $\mathbb{R}$ represents the set of  real numbers. $(\cdot)^{\T}$ denotes the vector/matrix transpose. The identity matrix of  is denoted by $\mbI$ and the trace of a matrix is denoted by $\operatorname{Tr}(.)$.  

\section{Problem  formulation}
We begin by the long-standing linear inference  problem formulation in which $m$
statistically independent signals are linearly mixed to yield  $l$ possibly noisy combinations,
\begin{equation}
	\label{eq::1}
	\mbx(t)=\mbA \mbs(t)+\mbn(t).
\end{equation} 
Let $\mbx(t)=[\mbx_1(t), \ldots, \mbx_l(t)]^{\T}$ denote the $l-$dimensional data vector made up of the mixture at time $t$ that is exposed to an additive measurement noise $\mbn(t)$.
Given no knowledge of the mixing matrix $\mbA\in \realR{m\times l}$, the goal is to recover the original source  signal vector $\mbs(t)=[\mbs_1(t), \ldots, \mbs_m(t)]^{\T}$ from the mixture. This problem is referred to as  blind source separation (BSS) in the  literature. A seminal work in this context is~\cite{cardoso1996equivariant} which suggests tuning and updating a separating matrix $\mbW \in \realR{l\times m}$ until the output 
\begin{equation} 
	\mby(t)=\mbW^{\T}\mbx(t),
\end{equation}
where $\mby(t)=[\mby_1(t), \ldots, \mby_m(t)]^{\T}$, is as close as possible to the  source signal vector of interest  $\mbs(t)$.
\subsection{Nonlinear Recursive Least Squares for  Blind Source Separation}
Assuming there exists a larger number of sensors than the  source signals, i.e. $l\geq m$, we can draw an analogy between  the BSS problem and the task of principal component analysis (PCA). In a sense, we are aiming to represent the random vector $\mbx(t)$ in a lower dimensional orthonormal subspace, represented by the columns of $\mbW$, as the orthonormal basis vectors.
By this analogy, both BSS and  PCA problems can be reduced to minimizing the objective function
\begin{equation}
	\label{loss1}
	\mathcal{L} (\mbW)=\mathbb{E}\left\{\|\mbx(t)-\mbW(\mbW^{\T}\mbx(t))\|_2^2\right\}.
\end{equation}
Assuming that $\mbx(t)$ is a zero-mean vector, it can be shown that the solution to the above  optimization problem is a matrix $\mbW$ whose columns are the $m$ dominant eigenvectors of  the data covariance matrix $\mathbf{C_x(t)}=\mathbb{E}\left\{ \mbx(t)\mbx(t)^{\T} \right\}$~\cite{romano2018unsupervised}. Therefore, the principal components which are the recovered source signals are mutually uncorrelated. As  discussed in~\cite{cardoso1996equivariant}, having uncorrelated data is not a sufficient  condition to achieve separation. In other words, the solutions to  PCA and  BSS do not coincide unless we address the  higher order statistics of the output signal $\mby(t)$. By  introducing nonlinearity into~\eqref{loss1},  we  will implicitly target higher order statistics of the signal~\cite{romano2018unsupervised}. This \emph{nonlinear PCA}, which is an extension of the  conventional PCA,  is made possible by considering the signal recovery objective:
\begin{equation}
	\label{nonlinear}
	\mathcal{L}(\mbW)=\mathbb{E}\left\{\|\mbx(t)-\mbW\mathbf{g}(\mbW^{\T}\mbx(t))\}\|_2^2\right\},
\end{equation} 
where $\mathbf{g}(\cdot)$ denotes an odd non-linear function applied element-wise on the vector argument. The proof of the connection between \eqref{nonlinear} and higher order statistics of  the source signals $\mbs(t)$ is discussed  in~\cite{pajunen1997least}.
While PCA is a fairly standardized
technique, nonlinear or robust PCA formulations  based on \eqref{nonlinear} tend to be multi-modal with several local optima---so they can be run from various initial points and possibly lead to  different ``solutions"~\cite{karhunen1997class}.
In~\cite{yang1995projection}, a recursive least squares algorithm for subspace estimation is proposed, which is further extended to the nonlinear PCA in~\cite{karhunen1997blind} for  solving the BSS problem. The algorithm in~\cite{karhunen1997blind} is useful as a baseline for  developing our  deep unfolded  framework for nonlinear PCA.

We consider a real-time and adaptive scenario in which, upon arrival of new data $\mbx(t)$, the  subspace of signal at time instant $t$  is recursively updated from the  subspace at time $t-1$ and the new  sample $\mbx(t)$~\cite{yang1995projection}. The separating matrix $\mbW$ introduced in~\eqref{nonlinear} is therefore replaced by $\mbW(t)$ and updated at each time instant $t$. The adaptive algorithm chosen for this task is the well-known recursive least squares (RLS)~\cite{Peng2023CVPR}. In the linear case, by replacing the  expectation in~\eqref{loss1} with a weighted sum,
we can attenuate the impact of the older samples, which is reasonable for a time-varying environment.
In this way, one can make sure the distant past will be forgotten and the resulting algorithm for minimizing~\eqref{loss1} can effectively track the statistical variations of the observed data.
By replacing $\mby(t)=\mbW(t)^{\T}\mbx(t)$ and  using an exponential weighting (governed by a \textit{forgetting factor}), the loss function in \eqref{loss1} boils down to:
\begin{equation}
	\label{loss2}
	\mathcal{L}(\mbW)=\sum_{i=1}^{t}\beta^{t-i}\|\mbx(i)-\mbW(t) \mby(i)\|^2,
\end{equation}
with the forgetting factor $\beta$ satisfying $0 \ll\beta \leq1$. Note that $\beta= 1$ yields the
ordinary method of least squares in which all samples are weighed equally while choosing relatively small  $\beta$ makes our estimation rather instantaneous, thus neglecting the past. Therefore, $\beta$ is  usually  chosen to be less than one, but also rather close to one for smooth tracking and filtering.

We can write the 
gradient of the loss function in \eqref{loss2} in its compact form  as 
\begin{equation}
	\label{grad}
	\nabla_{\mbW(t)} \mathcal{L} (\mbW)=-2\mathbf{C_{xy}}(t)+2\mathbf{C_{y}}(t)\mbW(t),
\end{equation}
where $\mathbf{C_{y}}(t)$ and $\mathbf{C_{xy}}(t)$ are the auto-correlation matrix of $\mby(t)$, 
\begin{equation}
	\label{cy}
	\mathbf{C_{y}}(t) =\sum_{i=1}^{t}\beta^{t-i}\mathbf y(i) \mathbf y(i)^{\T}=\beta \mathbf{C_{y}}(t-1)+\mby(t)\mby(t)^{\T},
\end{equation}
and the cross-correlation matrix  of $\mbx(t)$ and $\mby(t)$,
\begin{equation}
	\mathbf{C_{xy}}(t) =\sum_{i=1}^{t}\beta^{t-i}\mbx(i) \mby(i)^{\T}=\beta \mathbf{C_{xy}}(t-1)+\mbx(t) \mby(t) ^{\T},
\end{equation}
at the time instance $t$, respectively. Setting the gradient~\eqref{grad} to zero will result in the close-form separating matrix
\begin{equation}
\label{sol}	
\mbW(t)=\mathbf{C_{y}}^{-1}(t)\mathbf{C_{xy}}(t).
\end{equation}
A recursive computation of $\mbW(t)$ can be achieved using the RLS algorithm~\cite{haykin2014adaptive}. In RLS, the matrix inversion lemma enables a recursive computation of $\mbG(t)=\mathbf{C_{y}}^{-1}(t)$; see the derivations in Appendix. At each iteration of the RLS algorithm, 
$\mbG(t)$ is computed recursively as
\begin{equation}\label{rec}	
	\mbG(t)=\beta^{-1}\mathbf{G}(t-1)-
	\frac{\beta^{-2}\mathbf{G}(t-1)\mby(t) \mby(t)^{\T}\mathbf{G}(t-1)}
	{1+\beta^{-1}\mby(t)^{\T} \mathbf{G}(t-1)\mby(t)}.
\end{equation}
Consequently, the RLS algorithm provides the estimate $\mby(t)$ of the source signals. The steps of the RLS algorithm are summarized in Algorithm~\ref{algo1}. It appears that extending the application of RLS to the  nonlinear PCA  loss function  in \eqref{nonlinear} is rather straightforward. To accomplish this task, solely step 3 of Algorithm~\ref{algo1} should be modified to $\mby(t)=g(\mbW^{\T}(t-1)\mbx(t))$ in order to meet the nonlinear PCA criterion~\cite{yang1995projection}.
\begin{algorithm}[tb]
\caption{RLS Algorithm for Performing PCA}
\begin{algorithmic}[1]
\State\textbf{Initialize} $\mbW(0)$ and $\mbG(0)$
\For {$t=0,1,\ldots, T$}{}
\State{\indent} $\mby(t)=\mathbf{W}^{\T}(t-1)\mbx(t)$
\State{\indent}	$\mbh(t)=\mathbf{G}(t-1)\mby(t)$
\State{\indent}	$\mbf(t)=\frac{\mbh(t)}{\beta+\mathbf y(t)^{\T} \mbh(t)}$
\State{\indent}	$\mbG(t)=\beta^{-1}[\mathbf{G}(t-1)-\mbf(t) \mathbf h(t)^{\T}]$
\State{\indent}	$\mbe(t)=\mbx(t)-\mathbf{W}(t-1)\mby(t)$
\State{\indent}	$\mbW(t)=\mathbf{W}(t-1)+\mbe(t)\mbf(t)^{\T}$
\EndFor
\end{algorithmic}\label{algo1}
\end{algorithm}
 In order for  the RLS algorithm to optimize the  nonlinear PCA  loss function and converge to a separating matrix the choice of nonlinearity g(.) matters. We  refer to the analytical study presented in~\cite{karhunen1997class}   in which some  conditions beyond the oddity and  differentiability of the function  g(.) must be satisfied. 
For instance, $g(s)=s^3$ leads to an asymptotically stable  separation   only if the source  signals are positively kurtotic or super-Gaussian. Whereas if we choose a sigmoidal nonlinearity  $g(s)=tanh(\beta s)$ with $\beta >0$, then a sub-Gaussian density  is required for  the source signals  to be separated using RLS algorithm.

 In section~\ref{sec::deeprls}, we \emph{unroll} the iterations of the modified Algorithm~\ref{algo1}, for nonlinear PCA  onto the layers of a deep neural network where each layer resembles one iteration of the RLS algorithm.
 \subsection{Equivariant Adaptive Source Separation (EASI)}
In~\cite{cardoso1996equivariant}, the EASI algorithm is developed  by recurrent updates of the separating matrix as 
\begin{equation}
\label{eq::EASI}
\mbW(t+1)=\mbW(t)-\lambda_t \mbH(\mby(t))\mbW(t).
\end{equation}
Where $\lambda_t$ is a sequence of  positive step sizes and  $\mbH(.)$ is a matrix valued  function used to update the separating matrix. In~\cite{cardoso1996equivariant} this function is calculated as the relative gradient with respect to the objective function for blind source separation. The cross-cumulants of the source signals in $\mby(t)$ is  proposed as the objective function to be minimized as a measure of independence. $\mbH(.)$ is derived in~\cite{cardoso1996equivariant} as 
\begin{equation}\label{eq:grad-descent}
\mbH(\mby(t))=\mby(t)\mby(t)^{\T}-\mbI+g(\mby(t))\mby(t)^{\T}-\mby(t) g(\mby(t))^{\T}
\end{equation}
where l arbitrary nonlinear functions, $g_1(.),g_2(.),\ldots,g_l(.)$  are  used to define 
\begin{equation}\label{eq:nonlinear}
g(\mby(t))=[g_1(\mby_1(t)),\ldots,g_l(\mby_l(t))]^{\T}.
\end{equation}
The  choice of  this  nonlinear function is crucial to the performance of the  algorithm and is  dependent on the  distribution of sources. For instance, for sources with identical distibutions $g_i(.)=g(.)$ will be sufficient to perform seperation. In~\cite{cardoso1996equivariant}, it is illustrated that a  cubic nonlinearity  $g(s)=s^3$ leads to stability of separation in EASI algorithm only under the  constraint that  sum of kurtosis of any two source signals $s_i$ and $s_j$, $1\leq i,j\leq m$ are negative. $g(s)=tanh(s)$ is reported in~\cite{karhunen1997blind} to work satisfactorily for two sub-Gaussian sources using $\lambda_t >0$.
The nonlinear PCA algorithm require that the original source signals have a kurtosis with the same sign. Although this condition is somehow relaxed in the EASI
algorithm so that the sum of kurtosises for any pair of two sources must be negative, still some knowledge on the probability distribution of source signals is required to  choose the  nonlinearity. In the following sections, we propose to  learn  the nonlinearity in ~\eqref{eq:nonlinear} along with the  step size in~\eqref{eq::EASI} using deep unfolding networks. 

\section{The Proposed Framework}\label{sec::framework}
The estimation performance of the algorithms discussed above depends on a number of factors such as condition of mixing matrix $\mbA$, source signals, the step size parameter(s) and nonlinearity $g(.)$. We propose to  overparameterize the  algorithms so that we can determine the optimal stepsize and  proper nonlinearities as apposed to using fixed parameters.
In this section, we present the proposed deep  architectures  
\textit{Deep RLS} in~\ref{sec::deeprls} and \textit{Deep EASI} in~\ref{sec::deepeasi}.
The training procedure for these two architectures and the  SURE based  loss  function  will be  introduced in~\ref{sec::training} and \ref{sec::stein}, respectively.
\subsection{Deep RLS}\label{sec::deeprls}
As shown in~\cite{karhunen1997blind}, when applied to a linear mixture of  source signals (i.e., the BSS problem), the RLS algorithm usually approximates the true source signals well and successfully separates them. However, the number of iterations needed to converge may vary greatly depending on the initial values $\mbW(0), \mbG(0)$ and the forgetting parameter $\beta$. We introduce \textit{Deep RLS}, our deep unrolling-based framework which is designed based on the modified iterations of the algorithm~\ref{algo1}. More precisely, the  dynamics of the
$k$-th layer of \textit{Deep RLS} are given as: 
\begin{align}
		\mby(k)& =\mbg_{_{\nu_k}}(\mbW^{\T}(k-1)\mbx(k)), \label{deeprls1}\\
		\mbh(k)&=\mbG(k-1)\mby(k), \label{deeprls3}	\\
		\mbf(k)&=\frac{\mbh(k)}{\mathbf \omega_{k}+\mby(k)^{\T}\mathbf h(k)},\label{deeprls2} \\
		\mbG(k)&=\omega_{k}^{-1}[\mbG(k-1)-\mathbf{f}(k)\mbh(k)^{\T}],\label{deeprls4}\\
		\mbe(k)&=\mbx(k)-\mbW(k-1)\mby(k), \label{deeprls5}	\\
		\mbW(k)&=\mbW(k-1)+\mbe(k)\mbf(k)^{\T},\label{deeprls6} 
\end{align}
where $\mathbf x(k)$ is the  data vector at time instance $k$. The nonlinearity  $\mathbf g(\cdot)$  in the original RLS algorithm, which was chosen accroding to the  distribution of the source signals, is overparameterized  to $\mathbf g_{_{\nu_k}}(\cdot)$. Considering that neural networks with at least one hidden layer are universal approximators and they can be trained to approximate any mapping, we use a set of fully connected layers as  $\mathbf g_{_{\nu_k}}(\cdot)$. Weights and biases of these layers are represented by the learnable  parameter $\nu_k$ and $\omega_k\in\mathds{R}$ represents the trainable forgetting parameter.

Given $T$ samples of the data vector $\mbx(t)$,
our goal is to optimize the parameters $\bGamma$ of the network, where 
\begin{equation}
\label{weights}
\bGamma= \{\nu_k,\mathbf{\omega}_{k}\}_{k=1}^{T}.
\end{equation}
The output of the $k$-th layer, i.e. $\mathbf y(k)$, in~\eqref{deeprls1} is an approximation of the  source signals at the time instance $k$. 

\subsection{Deep EASI}\label{sec::deepeasi}
We consider the EASI iterations defined in~\eqref{eq::EASI} as  our baseline to reconstruct the  unfolded network. We over parameterize the iterations by introducing a learnable step-size $\lambda_t$ and  $\mbH_{\phi_t}$ for each layer $t$ as
\begin{equation}
\label{eq::deepeasi}
\mbW(t+1)=\mbW(t)-\lambda_t \mbH_{\phi_t}(\mby(t))\mbW(t)
\end{equation}
and
\begin{equation}
\mbH_{\phi_t}(\mby(t))=\mby(t)\mby(t)^{\T}-\mbI+g_{\phi_t}(\mby(t))\mby(t)^{\T}-\mby(t) g_{\phi_t}(\mby(t))^{\T},
\end{equation}
where  $ g_{\phi_t}(.)$ is realized using a few layers of  fully connected layers, with weights $\phi_t$, deployed  to approximate the best nonlinearity for  separation.  The   trainable parameters of  the network will be 
\begin{equation}
\label{eq::deepeasiparam}
\btheta=\{\phi_t,\lambda_t\}_{t=1}^{T}
\end{equation}
The output of the $t$-th layer, i.e. $\mathbf y(t)$ is 
\begin{equation}
	\label{eq::IO}
	\mby(t)=\mbW(t)^{\T}\mbx(t),
\end{equation}
\nocite{ajirlou2021machine}
\subsection{Training Procedure}
\label{sec::training}
An earlier version of this work proposed in~\cite{esmaeilbeig2020deep} and was successful in  recovering the source signals. However, it could only use a few number of  inputs $\mbx(t)$ because deeper networks with huge number of parameters were not feasible to be trained. Parameter sharing is a  technique in  deep learning which regularizes the  network to avoid this problem. Parameter sharing makes it possible to extend and apply the model to signals of different lengths and generalize across them.  In \emph{Deep RLS} architecture proposed in~\cite{esmaeilbeig2020deep}, we designed a  multi-layer feedforward neural network in which we had separate learnable parameters for each time index and we could not generalize to sequence lengths not seen during training. Recurrent Neural Networks (RNN) are  introduced to overcome this limitation. Inspired by the architecture of   these neural networks and \textit{back-propagation through time (BPTT)} as their training process, we propose the following  loss function for training   the proposed  unrolled algortihms.
An RNN  maps an input sequence to an output sequence of the same length. The total loss for a given sequence of $\mbx(t)$ values paired with a sequence of $\mby(t)$ values is the sum of the losses over all the time steps. For example, if $L(t)$ is the mean squared error (MSE) of reconstructing $\mbs(t)$ given $\mby(t)$ then
\begin{equation}
	\label{lt}
	\mathcal{L}(\mathbf{s}, \mathbf{y}) = \sum_{t=1}^T L{(t)} = \sum_{t=1}^T \|\mbs(t)-\mby(t)\|^2_2,
\end{equation}
where $\mby=[\mby(1), \ldots, \mby(T)]^{\T}$ and $\mbs=[\mbs(1), \ldots, \mbs(T)]^{\T}$.
In order to apply BPTT, the gradient of the loss function $L(t)$ with respect to the trainable parameters is required.
This is challenging to do by hand, but made easy by  the auto-differentiation capabilities is PyTorch~\cite{NEURIPS2019_9015}, which is used throughout our experiments in section~\ref{sec::numerical}.

While training  \textit{Deep RLS}, one needs to consider the constraint that the forgetting parameter must satisfy $0<
\beta\leq 1$. Hence, in order to impose such a constraint, one can regularize the loss function ensuring that the network chooses proper weights $\{\omega_{k}\}_{k=1}^T$ corresponding to a feasible forgetting parameter at each layer. Accordingly, we define the loss function used for training  the proposed architecture as 
\begin{equation} \label{eq::eq1}
	\begin{split}
		\mathcal{L}(\mbs,\mby,\Gamma) & = \sum_{t=1}^T \|\mbs(t)-\mby(t)\|^2_2+\\
		& \underbrace{	\lambda \sum_{t=1}^{T}
			\mathrm{ReLU}(-\mathbf{\omega}_{t})
			+\lambda \sum_{t=1}^{T} \mathrm{ReLU}(\mathbf{\omega}_{t}-1) }_\text{regularization term for the forgetting parameter},
	\end{split}
\end{equation}
where  $\mathrm{ReLU}(\cdot)$ is  the well-known Rectifier
Linear Unit function extensively used in the deep learning literature. For  training the  \textit{Deep RLS} (or \textit{Deep EASI}) network we employed the  training  process in Algorithm~\ref{algo2}.\\
\begin{algorithm}[t]
\caption{Training Procedure for \textit{Deep RLS(or Deep EASI)}.}
	\begin{algorithmic}[1]
		\State \textbf{Initialize} $\mbW(0)$ and $\mathbf \mbG(0)$
		\For {$\textrm{epoch}=1, \ldots, N$}{}{\For {$t=1, \ldots, T$}{}
			\State{\indent} Feed $\mathbf x(t)$ to the network 
				\State{\indent} Apply the recursion in~\eqref{deeprls1}-\eqref{deeprls6} (or \eqref{eq::deepeasi})
			\State{\indent} Compute the loss function in~\eqref{eq::eq1} (or~\eqref{lt})
			\State{\indent} Use BPTT  to update $\bGamma$ (or $\btheta$) \EndFor}\EndFor
	\end{algorithmic}\label{algo2}
\end{algorithm}

\subsection{Stein's Unbiased Risk Estimator (SURE)}\label{sec::stein}
In 1981 Charles Stein in~\cite{stein1981estimation} showed that for $\mbx=\bmu+\bn $ and $\mbn \sim \mathcal{N}(\bzero,\sigma^{2}\mbI)$, the SURE
statistic defined as
\begin{equation}
\text{SURE}(\hat \bmu,\mbx)=-l \sigma^2 +  \|\hat{\bmu}(\mbx)-\mbx\|^{2}_2 +2 \sigma^2 \nabla \hat{\bmu}(\mbx) 
\end{equation}
is an unbiased estimate of the $l_2$ risk of the  estimator $\hat{\bmu}$ in the sense that $\expecE{\text{SURE}}=\expecE{\|\hat{\bmu}(\mbx)-\bmu\|^{2}}$.  The  bottleneck  in 
 evaluation of  SURE is   evaluation of the  divergence $\nabla \hat{\bmu}(\mbx)=\sum \frac{\partial\hat{\bmu_i}(\mbx)}{\partial x_i}$.
 A generalization of this technique
known as generalized SURE was proposed in~\cite{eldar2008generalized} to estimate the MSE associated with estimates
of $\mbs$ from  linear measurements $\mbx = \mbA\mbs+ \mbn$, where 
$\mbA \neq \mbI$, and $\mbn$ has known covariance and
follows any distribution from the exponential family. For the estimators $f_{\btheta}(.)$ parameterized over $\btheta$  which receive the noisy observation $\mbx$ and provide an estimate of sources $\mbs$, expectation of   the generalized  SURE is~\cite{metzler2018unsupervised} 
\begin{align}\label{eq:GSURE1}
&\mathbb{E}\{\|\hat{\mbs}- \mbs\|_2^2 \}\nonumber\\
&=\mathbb{E}\{\|\mbP \mbs- \mbP f_{\btheta}(\mbx)\|_2^2 \}= 
\mathbb{E}\{\|\mbP \mbs\|_2^2+\|\mbP f_{\theta}(\mbx)\|_2^2\nonumber\\
&+2 \sigma ^2 \nabla(f_{\btheta}(\mbx))-2f_{\btheta}(\mbx)^{\T}\mbA^{\dagger}\mbx\},
\end{align}

where the orthonormal projection onto  the range space of  $\mbA$ is  represented by $\mbP=\mbA(\mbA^{\T}\mbA)^{-1}\mbA^{\T}$ and  $\mbA^{\dagger}$ is the pseudoinverse of $\mbA$. 

The last three terms in~\eqref{eq:GSURE1} are dependant on the  parameters of the estimator i.e. $\btheta$. Considering layer $t$ of  \textit{Deep EASI}  as an estimator of the source signal $\mbx(t)$,  we propose to train the network’s parameters $\btheta$ by incorporating the SURE loss at time $t$ as
\begin{equation}\label{GSURE_lt}
L(t)=\|\mbP f_{\btheta}(\mbx(t))\|^2+\\2 \sigma ^2 \nabla(f_{\btheta}(\mbx(t)))-2f_{\btheta}(\mbx(t))^{\T}\mbA^{\dagger}\mbx(t).
\end{equation}
In this  equation, the  \textit{Deep EASI}  network,  as an estimator for source signal at time instance $t$, is denoted by $f_{\btheta}(\mbx(t))$. Consequently, the  divergence is 
\begin{align}
\label{eq::div_def}
 \nabla(f_{\btheta}(\mbx(t)))	&= \sum_{i=1}^{l}\frac{\partial f_{\btheta}(\mbx(t))}{\partial \mbx_i}\nonumber\\
 &=\sum \frac{\partial}{\partial x_i}(\mbW(t)^{\T}\mbx(t))\nonumber\\
&=\operatorname{Tr}(\mbW(t)).
\end{align}
 By Substituting~\eqref{eq::div_def}  in \eqref{GSURE_lt}, the SURE  loss function for
training the\textit{ \textit{Deep EASI}} network is 
\begin{align}
 & \mathcal{L}(\mathbf{A}, \mathbf{x})= \sum_{t=1}^T L{(t)}\\
 &=  \sum_{t=1}^T  \|\mbP f_{\btheta}(\mbx(t))\|^2+2 \sigma ^2 \operatorname{Tr}(\mbW(t))-2f_{\btheta}(\mbx(t))^{\T}\mbA^{\dagger}\mbx(t).\nonumber
\end{align}

It is worth mentioning that as discussed in~\cite{nobel2022tractable}, the  SURE loss function can not be analytically  derived  for all  estimators but it can be tractably evaluated using the methods introduced in~\cite{nobel2022tractable}. We leave derivation of this  loss  for   \textit{deep RLS} for future extension of this study.

While training   \textit{Deep EASI} and  \textit{Deep RLS}, we synthetically produce the   observations $x(t)$ for $t=1,\ldots,T$ to use as the training set. Therefore, we have access to  the sources $\mbs(t)$, the mixing matrix $\mbA$  and the variance of noise $\sigma^{2}$. Therefore evaluation of Evaluating the SURE loss in~\eqref{GSURE_lt}  is  possible  while training the network. Backpropagation on the SURE Loss function is feasible by means of PyTorch’s auto-differentiation capabilities, which is used throughout much of our experiments below.

The blindness of  our method refers to the test in which we only observe the mixtures. In our numerical experiments, we will deploy the  SURE loss function in~\eqref{GSURE_lt} for  training the  \textit{Deep EASI} network and illustrate the  performance improvement  over the initially proposed loss function in~\eqref{lt}. 
Evaluating the SURE loss in~\eqref{GSURE_lt}  does not require the ground truth $\mbs(t)$ and therefore the  learning procedure will be unsupervised.

\section{Numerical  Study}\label{sec::numerical}
In this section, we demonstrate the performance of the proposed \textit{Deep RLS} and  \textit{Deep EASI} algorithms.
The proposed framework was implemented using the PyTorch library~\cite{NEURIPS2019_9015} and the Adam stochastic optimizer~\cite{kingma2014adam} with a constant learning rate of $10^{-4}$ for training purposes.
The training was performed based on the data generated via the following model. For each time interval $t=0,1,...,T$,  elements of the vector
$\mbs(t)$ is generated from a sub-Gaussian distribution. For data generation purposes, we have assumed the source signals to be i.i.d. and uniformly distributed, i.e.,
$\mbs(t)\sim \mathcal{U}(0,\bone)$. The mixing matrix $\mathbf{A}$ is assumed to be
generated according to $\mathbf A \sim \mathcal{N}(\mathbf 0, \mathbf I)$. For each sample data, a new mixing matrix $
\mbA$ is generated i.e. there is not any two samples in the training and test sets that have the same mixing matrix.

 We performed  the training of the proposed \textit{Deep RLS} and  \textit{Deep EASI} networks using the batch learning process with a batch size of 40 and trained the network for $N=100$ epochs.
A training set of size $10^3$ and test set of size $10^2$ was chosen. In this section we study performance of our proposed source  separation algorithms in test simulations where the mixing matrix and source signals and  noise variance are known. Therefore, the  network will be trained in a supervised manner and also  the  mixing matrix  and variance of  the  samples in the  training set are used to train regularized network with  SURE loss  function in section~\ref{sec::stein}.

The quantitative measure  to evaluate  the   performance of the networks is  the average of the mean squared error (MSE) defined as $\mathrm{MSE} =(1/T)\sum_{t=1}^{T}\| \mathbf \mbs(t)-\mathbf \mby(t)\|^2_2$.

\begin{figure}[t]
\centering 
	\includegraphics[width=0.9\columnwidth]{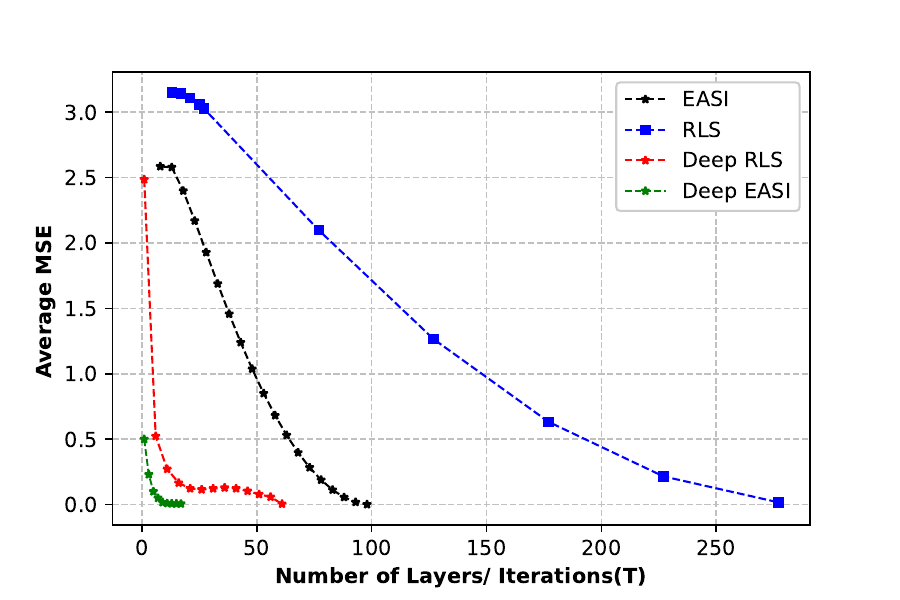}
	\caption{\small The average MSE of recovering $m=3$ source signals from $l=3$ observations using the \textit{Deep-RLS} network, RLS~\cite{karhunen1997blind} with $\beta=0.99$, EASI~\cite{cardoso1996equivariant}  and 
		\label{fig::1}	
		\textit{Deep-EASI} vs. the number of layers/iterations $T$, when trained by MSE loss for $N=100$ epochs with a learning rate of $10^{-4}$.}
	
\end{figure}

In Fig.~\ref{fig::1}, we  demonstrate the performance of the  proposed \textit{Deep RLS} algorithm, \textit{Deep EASI} algorithm  and   the base-line RLS~\cite{karhunen1997blind} and EASI~\cite{cardoso1996equivariant} algorithms (where no parameter is learned) in terms of $\mathrm{MSE}$ versus the number of  time samples $T$.
Observing Fig.~\ref{fig::1}, one can deduce that owing to its hybrid nature, the proposed \textit{Deep EASI} methodology  significantly outperforms its  counterparts in terms of average MSE and converges with far less iterations. In particular, we can observe that \textit{Deep EASI} and \textit{Deep RLS}  achieve a very low average MSE with as few as $50$ iterations, while the EASI and RLS algorithms require at least $100$ and $250$ iterations to converge, respectively. Accordingly, this results demonstrate the effectiveness of the learned parameters.

\begin{figure}[h]
\centering 
\includegraphics[width=0.9\columnwidth]{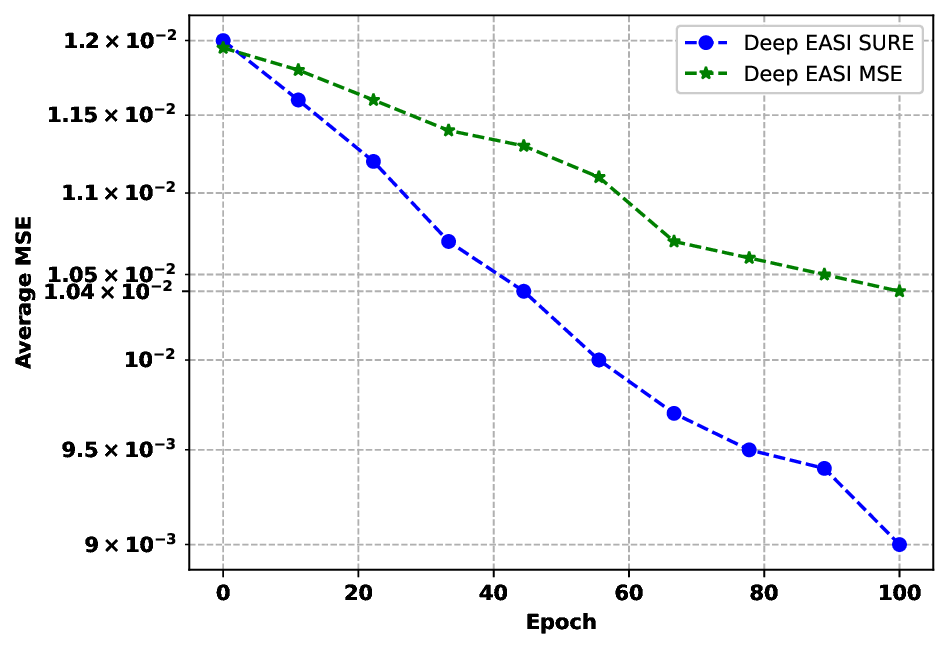}
\caption{\small The performance of \textit{Deep EASI} trained with  MSE in~\eqref{lt}  in comparison with  SURE loss function in~\eqref{GSURE_lt}
.}
\label{fig::2}
\end{figure}

Fig.~\ref{fig::2} illustrates the average $\mathrm{MSE}$ on test set  for \textit{Deep EASI} network trained using two different loss functions. The efficacy of  SURE-based training in comparison with MSE loss is evident in every epoch of the training.

\section{Conclusion}\label{sec:conclusion}
In this paper, we  introduced two  deep unrolling-based  frameworks for  adaptive filtering and demonstrated that  the   unrolled  networks, trained as recursive neural networks,  outperform their baseline counterparts. 
Moreover, we incorporated 
Stein's unbiased risk estimator as  a surrogate loss  function for training the  deep  architectures which introduced further improvement in estimating the  source signals.
\section{Appendix: The RLS Recursive Formula}
\label{sec:app}
Let $\mathbf A$, $\mathbf B$, and $\mathbf D$ be positive definite matrices such that
$
 	\mathbf A=\mathbf B^{-1}+\mathbf{cD}^{-1}\mathbf{c}^{\T}.
$ 
Using the matrix inversion lemma, the inverse of $\mathbf A$ can be expressed as
\begin{equation}
	\mathbf A^{-1}=\mathbf B-\mathbf{Bc}(\mathbf{D}+\mathbf{c}^T\mathbf{Bc})^{-1}\mathbf c^T\mathbf B. 
\end{equation}
Now, assuming  that the auto-correlation matrix $\mathbf{C_y}(t)$ is  positive definite (and thus non-singular), by choosing $\mathbf{A=C_y}(t)$,  $\mathbf{B}^{-1}=\beta\mathbf{C_y}(t-1)$, $\mathbf{c}=\mby(t), \mathbf D^{-1}=\mbI$, one  can compute $\mbG(t)=\mathbf{C_y}^{-1}(t)$ as proposed in \eqref{rec}.
\section{Acknowledgement}
The  authors would  like to express gratitude to Dr. Shahin Khobahi for his  help with developing  the  ideas in a preliminary version of this manuscript, currently available on arXiv~\cite{esmaeilbeig2020deep}.
\bibliographystyle{IEEEtran}
\bibliography{refs}

\end{document}